# Optofluidic random laser

**Shivakiran Bhaktha B.N.,**†*[a] **Xavier Noblin**[a] **and Patrick Sebbah***[b]

[a] *Laboratoire de Physique de la Matière Condensée, CNRS UMR 7336 and Université de Nice-Sophia Antipolis, Parc Valrose, 06108, Nice Cedex 02, France.*
[b] *Institut Langevin, ESPCI ParisTech, CNRS 7587, 10 Rue Vauquelin, 75005 Paris, France.*
† *Presently at Department of Physics and Meteorology, Indian Institute of Technology Kharagpur, Kharagpur, 721 302, India.*
*kiranbhaktha@phy.iitkgp.ernet.in, *patrick.sebbah@espci.fr

**An active disordered medium able to lase is called a random laser (RL). We demonstrate random lasing due to inherent disorder in a dye circulated structured microfluidic channel. We consistently observe RL modes which are varied by changing the pumping conditions. Potential applications for on-chip sources and sensors are discussed.**

Random laser emission arises in mirrorless scattering active media when multiple scattering within the gain allows overcoming the losses.[1] Coherent laser emission has been reported in various random active media.[2-5] Increasing interest in this field in the recent years is due to the ease of fabrication of random lasers (RLs), together with their unique properties.[6-8]

The combination of optical devices with microfluidics has led to the development of the field of optofluidics. The properties of optical components such as optical cavities or lasers can be dynamically controlled with microfluidic systems, a versatility which is not readily available with solid-state optical components.[9] Light can be manipulated at the micro scale, forming attractive systems for lab-on-a-chip applications[10] and opening new avenues for sensing applications. Optofluidic lasers proposed to date are based on dye-injected microchannels wherein optical feedback is provided by various mechanisms such as an external cavity,[11] a distributed grating,[12] or whispering gallery modes.[13] Recently, an attempt to use scattering as the feedback mechanism has been proposed by introducing $TiO_2$ nanoparticles in the dye flow.[14] However, lasing was found to be more difficult to reach in the presence of colloidal scatterers. Moreover, non-reproductive spectra resulted from the continuous change of the flowing scattering medium.

In this paper, we structure a polydimethylsiloxane (PDMS) microfluidic channel and fill it with an ethanolic dye solution. Spontaneous emission from the dye molecules is stimulated along the microfluidic channel by a stripe-shaped pump and multiply scattered by the inherent randomness of the structure. When the gain overcomes the losses, efficient low-threshold random lasing is observed. Because the random structure is static, the lasing modes are stable but sensitive to any local change of the structure. We further demonstrate variation of RL modes by changing parameters like the pump position or the refractive index and show the extreme sensitivity of this device to local perturbations. Sensor capabilities are demonstrated leading to potential unique applications.

The 3-mm long PDMS microfluidic channel was fabricated following the protocol described by Xia *et al.*[15] A 10:1 PDMS:cross-linker mixture (Sylgard 184) was poured onto a 28 µm-thick negative photoresist SU-8 mold which had the desired structure imprinted on it. The PDMS mixture was then degassed for 10 min at a few mmHg vacuum pressure, and cured at 90°C for 1hr 30'. After creating holes in the device for the inlets, the channel was bonded on glass slides by plasma treatment. Rhodamine 6G dye solution with $2.5 \times 10^{-3}$ M concentration in ethanol is circulated into the microchannel through the tubes connected to the inlets. The dye flow is particularly useful here since it allows dye regeneration and prevents its bleaching.

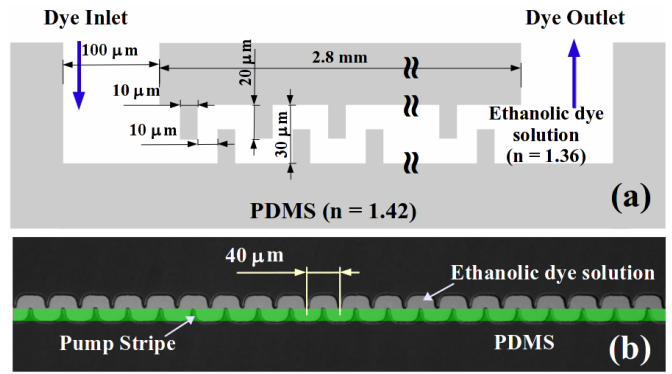

**Fig. 1** (a) Schematic view of the 3-mm long microfluidic channel with 10 µm-thick PDMS walls positioned periodically along its length with a 40 µm period. (b) An optical microscope image of the dye filled channel obtained at 10× magnification. The green stripe represents the pumping scheme used for performing RL emission studies.

Fig. 1a shows the periodic rectangular structure of the mask imprinted onto the photoresist mold. The 40 µm-periodic design has 10 µm-thick, 20 µm-long vertical polymer pegs, positioned 30 µm apart, along both sides of the microfluidic channel. An optical microscope image at 10× magnification of a portion of the structured channel, 28 µm-deep, defined by the thickness of the mold, is shown in Fig. 1b. Photons propagating along the green line shown in Fig.1b see therefore a one-dimensional layered medium of alternating dielectric and dye layers. Random fabrication fluctuations result in a tolerance of ± 0.65 µm in the periodicity of the structure, which is of the order of the wavelength of the emitted photon. At the optical scale, the structure is therefore completely random for light and not periodic as it may appear to the naked eye in Fig.1b.

The second harmonic of a pulsed Nd:YAG laser (with 6 ns pulsewidth, 20 Hz repetition rate, and λ=532 nm) is used to pump the dye circulating along the channel. The pump beam is shaped into a 3mm-long, 10 µm-thick stripe by a $f$ = 50 mm cylindrical lens and covers half the width of the channel, as depicted in Fig. 1b. This narrow-strip pump provides uniform illumination and forces the dye emission along the length of the channel. The emission spectra are recorded with the fiber probe HR4000

(Ocean Optics) spectrometer having a peak-to-peak spectral resolution of 0.11 nm. During the experiments, the random optofluidic channel is imaged with the help of a Zeiss Axioexaminer microscope and a Hamamatsu Orca-R2 silicon CCD camera, to ensure perfect alignment of the pump-stripe with the channel.

partial pumping scheme is depicted in Fig. 3a. A stripe length of 300 μm was chosen, and scanned along the length of the channel by varying the distance 'd' between the center of the pump stripe and one end of the channel. Different pump regions correspond to different random configurations, resulting in a change in the RL emission spectrum. This is illustrated in Fig. 3b for three different partial pump positions, $d$ = 0.70 mm, 1.20 mm, and 2.55 mm, wherein the RL emission peaks are observed at 562.5 nm, 563.6 nm, and 561.5 nm respectively. Thus, by altering the position and length of partial pump, different set of lasing modes can be activated.

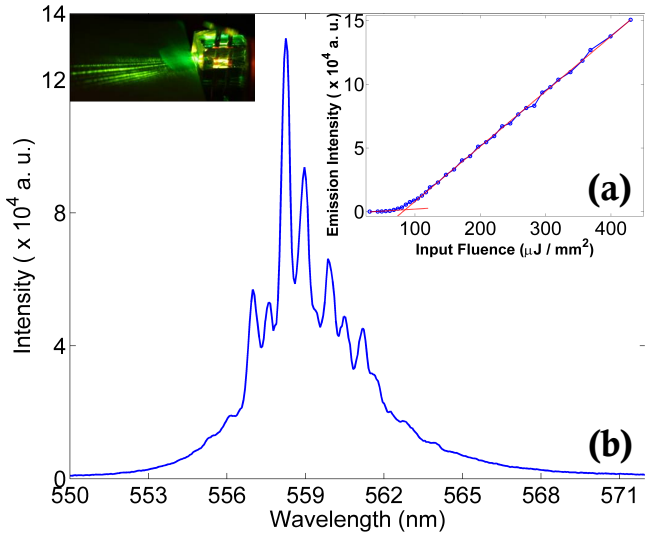

**Fig. 2** (a) The emission intensity of the optofluidic RL is plotted against the input fluence, and the threshold of lasing is determined to be about 80 μJ/mm². (b) Emission spectrum of the RL with its modes appearing at random spectral positions, recorded at a pump fluence above the lasing threshold at 233 μJ/mm². Inset to the left is a photograph of the emission from the dye filled optofluidic channel when pumped by a green laser.

Due to the inherent disorder of the structure, the stimulated photons, channelized by the pump stripe, are multiply scattered at each PDMS-dye interface, and eventually exhibit RL action when the losses are overcome. Typical data of the RL action are shown in Fig. 2. Fig. 2a is a plot of integrated emission intensity versus input fluence. A threshold characteristic typical of a laser is observed and a threshold value of about 80 μJ/mm² is found by linear fit. We point out that this value of lasing threshold is similar to values found for precisely designed optofluidic lasers.[11,13]

The randomly positioned sharp peaks that appear over the broad emission spectrum in Fig. 2b correspond to various modes of the RL. The typical linewidth of the modes is ~ 0.3 nm. The spectral positions of the modes are consistent and do not vary from shot to shot, in contrast to other RLs.[14,16] The inset of Fig. 2a shows the dye-filled optofluidic device being pumped by the green laser and the intense yellow emission of the dye. The photoluminescence of the dye is isotropic, whereas the RL emission is directional and diffracted by the on-average periodic structure.

Subsequently, we demonstrate spectral sensitivity of the laser emission on the pump region by changing the stripe position and length, and probe different parts of the random system. The

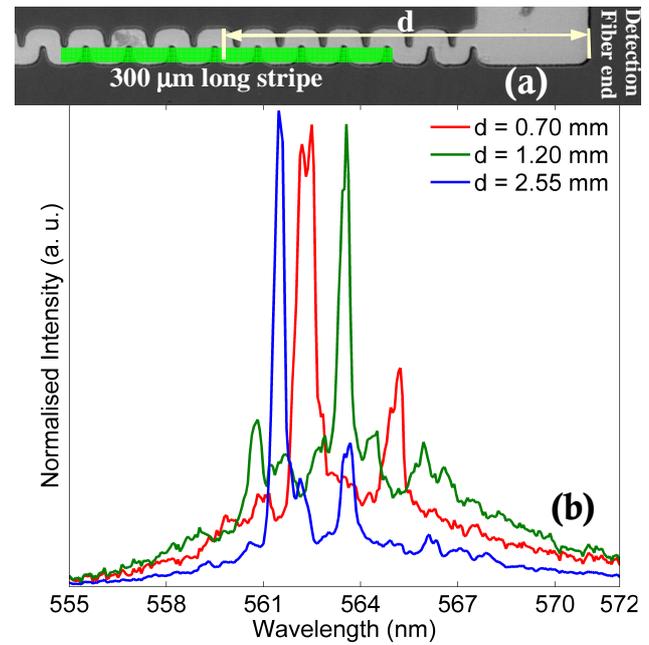

**Fig. 3** (a) A 300 μm long pump stripe is translated along the length of the channel, varying $d$, to study the spectral sensitivity of the random lasing modes to the pump region. The spectra recorded for three different values of $d$ are plotted in (b).

Finally, we probe the sensitivity of the optofluidic random laser to a local perturbation. It is well known that the modes of a scattering random medium are extremely sensitive to a local change of the disorder, either in position or in the optical characteristics of the medium.[17,18] Here, we measure the impact on the laser emission of a change of the refractive index in the vicinity of the channel. We designed a modified structure composed of two parallel channels similar to the previous one (Fig. 1) as shown in Fig. 4a. The average period of both channels is 120 μm with a standard deviation of ± 0.84 μm. The system is carefully designed to align alternately the short horizontal sections of each channel. Consequently, the 800 μm-long laser stripe covers both channels as shown in the Fig. 4a. One channel (bottom in Fig. 4a) is circulated with the dye solution, while the other (top in Fig. 4a) is initially filled with air. Introducing pure ethanol solution in this channel changes progressively the index of refraction seen by the stimulated photons. While the ethanol flow is away from the pumping region, emission spectrum

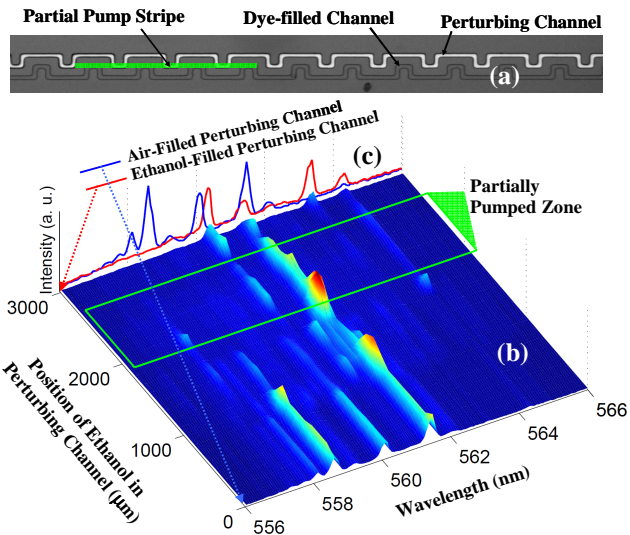

**Fig. 4** (a) An optical microscope image of the dye filled and perturbing channel is shown along with the employed pump scheme; (b) shows the effect of a local refractive index perturbation on the RL modes when the ethanol in the perturbing channel approaches the 800 μm long partially pumped zone marked by the green rectangle in the plot. The spectra for the case of fully-air-filled and fully-ethanol-filled perturbing channel are shown in (c).

remains essentially unchanged, whereas new lasing modes replace the original ones as the flow reaches the pumping region as seen in the 3-dimensional plot in Fig. 4b. The modes at 558.1 nm, 558.6 nm and 560.1 nm stop lasing when the pumping region (green area) is reached by the ethanol, while modes at 560.3 nm, 562.5 nm, 563.4 nm, and 564.3 nm start to lase. Globally, the emission spectrum is shifted toward longer wavelength to compensate for the increased index of refraction within the laser cavity, as in a standard Fabry-Perrot-type laser, which can be observed in the two spectra in Fig. 4c, for the case of fully-air-filled and fully-ethanol-filled perturbing channel. The emergence of different laser lines at random wavelengths is however a specificity of the random laser. Reproducing this experiment while pumping a different area of the microfluidic channel will give a similar red shift of the emission spectrum but different emerging lasing modes. This spatial dependence of the laser signature of a local perturbation should be of strong interest in the design of a multiplexed sensor.

In conclusion, we have designed and fabricated a structured optofluidic microchannel with inherent disorder. When filled with an ethanolic dye solution and pumped by a laser stripe mirrorless RL action is observed. The release of precise fabrication constraints, required for conventional lasers, makes the optofluidic RL a promising source for lab-on-a-chip applications. We further illustrate partial pumping mechanism as a means to vary the RL emission. Finally, we demonstrate this type of device to be a potential original on-chip sensor based on sensitivity of the emission spectrum to a local change of the refractive index.

The authors wish to acknowledge Amélie Trichon for the help in sample preparation. This work was supported by the ANR under Grant No. ANR-08-BLAN-0302-01 and the Groupement de Recherche 3219 MesoImage.

## Notes and references

1. V. S. Letokhov, *Sov. Phys. JETP*, 1968, **26**, 835.
2. N. M. Lawandy, R. M. Balachandran, A. S. L. Gomes and E. Sauvain, *Nature*, 1994, **368**, 436.
3. M. A. Noginov, H. J. Caulfield, N. E. Noginova and P. Venkateswarlu, *Opt. Commun.*, 1995, **118**, 430.
4. C. Gouedard, D. Husson, C. Sauteret, F. Auzel and A. Migus, *J. Opt. Soc. Am. B*, 1993, **10**, 2358.
5. H. Cao, Y. G. Zhao, S. T. Ho, E. W. Seelig, Q. H. Wang and R. P. H. Chang, *Phys. Rev. Lett.*, 1999, **82**, 2278.
6. D. S. Wiersma, *Nat. Phys.*, 2008, **4**, 359.
7. W. L. Sha, C.-H. Liu and R. R. Alfano, *Opt. Lett.*, 1994, **19**, 1922.
8. J. Andreasen, A. Asatryan, L. Botten, M. Byrne, H. Cao, L. Ge, L. Labonté, P. Sebbah, A. D. Stone, H. E. Türeci and C. Vanneste, *Adv. Opt. Photon.*, 2011, **3**, 88.
9. Zhenyu Li and Demetri Psaltis, *Microfluid Nanofluid*, 2008, **4**, 145.
10. C. Monat, P. Domachuk and B. J. Eggleton, *Nature Photonics*, 2007, **1**, 106.
11. B. Helbo, A. Kristensen and A. Menon, *J. Micromech. Microeng.*, 2003, **13**, 307.
12. W. Song, A. E. Vasdekis, Z. Li and D. Psaltis, *Appl. Phys. Lett.*, 2004, **94**, 051117.
13. S. K. Y. Tang, Z. Li, A. R. Abate, J. J. Agresti, D. A. Weitz, D. Psaltis and G. M. Whitesides, *Lab Chip*, 2009, **9**, 2767.
14. K. C. Vishnubhatla, J. Clark, G. Lanzani, R. Ramponi, R. Osellame and T. Virgili, *Appl. Opt.*, 2009, **48**, G114.
15. Y. N. Xia, G. M. Whitesides, *Annu. Rev. Mater. Sci.*, 1998, **28**, 153.
16. S. Mujumdar, M. Ricci, R. Torre and D. S. Wiersma, *Phys. Rev. Lett.*, 2004, **93**, 053903.
17. K. Y. Bliokh, Y. P. Bliokh, V. Freilikher, A. Z. Genack and P. Sebbah, *Phys. Rev. Lett.*, 2008, **101**, 133901.
18. L. Labontée, C. Vanneste and P. Sebbah, *Accepted for publication in Opt. Lett.*